\documentclass[]{spie}  %>>> use for US letter paper
\usepackage{amsmath,amsfonts,amssymb}
\usepackage{graphicx}
\usepackage[colorlinks=true, allcolors=blue]{hyperref}

\title{Fast modulation and dithering for the NFIRAOS Pyramid Wavefront Sensor}

\author[a]{Edward L. Chapin$^*$}
\author[a]{David Andersen}
\author[b]{Owen Brown}
\author[a]{Jeffrey Crane}
\author[a]{Adam Densmore}
\author[a]{Jennifer Dunn}
\author[a]{Tim Hardy}
\author[a]{Glen Herriot}
\author[a]{Dan Kerley}
\author[a]{Olivier Lardiere}
\author[a]{Jean-Pierre V\'eran}

\affil[a]{National Research Council Herzberg, 5071 W Saanich Rd, Victoria, V9E 2E7, Canada}
\affil[b]{University of Victoria, Victoria, Canada}

\authorinfo{$^*$ed.chapin@nrc-cnrc.gc.ca, phone +1 250 363 6971}

\begin{document}
\maketitle

\begin{abstract}
  The Narrow Field InfraRed Adaptive Optics System (NFIRAOS) for the Thirty Meter Telescope (TMT) will use a natural guide star (NGS) Pyramid Wavefront Sensor (PWFS). A 32-mm diameter Fast Steering Mirror (FSM) is used to modulate the position of the NGS image around the tip of the pyramid. The mirror traces out a circular tip/tilt pattern at up to 800 Hz (the maximum operating frequency of NFIRAOS), with a diameter chosen to balance sensitivity and dynamic range. A circular dither pattern at 1/4 the modulation frequency is superimposed to facilitate optical gain measurements. The timing of this motion is synchronized precisely with individual exposures on the PWFS detector, and must also be phased with other wavefront sensors, such as Laser Guide Star Wavefront Sensors (LGSWFS) and the On-Instrument Wavefront Sensors (OIWFS) of NFIRAOS client instruments (depending on the observing mode), to minimize latency. During trade studies it was decided to pursue a piezo actuator from Physik Instrumente (PI) using a monocrystalline piezo material, as more conventional polycrystalline devices would not meet the lifetime, stroke, and frequency requirements. Furthermore, PI claims excellent stability and hysteresis with similar piezo stages, rendering sensor feedback unnecessary. To characterize the performance of this mechanism, and to verify that it can function acceptably in open-loop, we have operated the stage on a test bench using a laser and high-speed position sensing devices (PSDs) both at room temperature and at the cold -30\,C operating temperature of NFIRAOS. We have also prototyped the software and hardware triggering strategy that will be used to synchronize the FSM with the rest of NFIRAOS.
\end{abstract}


%
% INTRODUCTION
%

\section{INTRODUCTION}
\label{sec:intro}

\begin{figure}[ht]
  \begin{center}
    \includegraphics[width=0.6\linewidth]{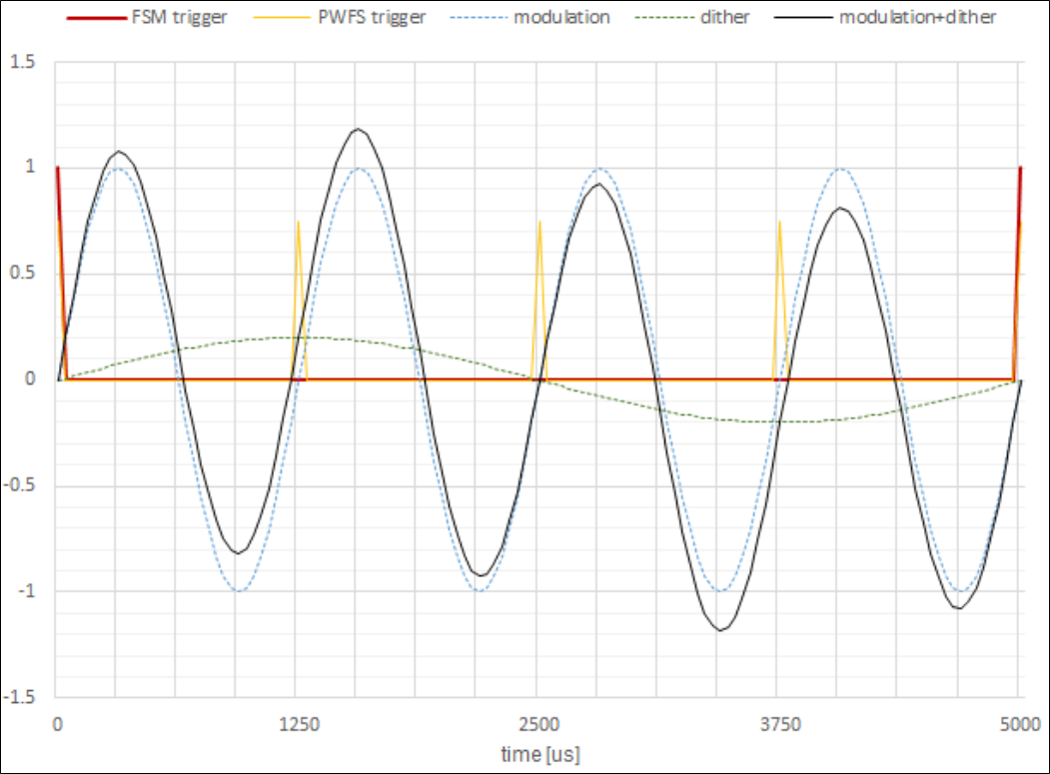}
  \end{center}
  \caption{\label{fig:waveforms}
  Illustration of the combined modulation and dither signals for one of the two FSM axes, along with triggers indicating when the FSM waveform repeats, and when PWFS exposures occur.
}
\end{figure}

The Narrow Field InfraRed Adaptive Optics System (NFIRAOS) will be the first-light adaptive optics (AO) system for the Thirty Meter Telescope (TMT)\cite{crane2018}. Wavefront sensing will be provided by 6 Laser Guide Star (LGS) wavefront sensors (WFS), a natural guide star (NGS) Pyramid Wavefront Sensor (PWFS), and also low-order NGS On-Instrument Wavefront Sensors (OIWFS) and On-[Science-]Detector Guide Windows (ODGW) provided by the three client instruments. The LGS WFS and PWFS sense visible light, while near-infrared light is passed downstream to the client instruments via a beam splitter. The PWFS will incorporate a 32-mm diameter Fast Steering Mirror (FSM) to modulate the position of the NGS on the tip of the pyramid. This mirror traces out a circular tip/tilt pattern at up to 800 Hz (the maximum operating frequency of NFIRAOS), with a diameter chosen to balance sensitivity and dynamic range. In addition, a slower, low-amplitude circular dither pattern at 1/4 the modulation frequency is superimposed to facilitate optical gain measurements. This process uses synchronous detection to assess the apparent radius of the dither circle from the PWFS Tip/Tilt measurements during observing. The tip and tilt measurements are correlated with sines and cosines at the dither frequency, and then the four correlation signals are time-averaged and added in quadrature to obtain the radius. This radius is compared with the known motion, which is precisely calibrated in the daytime versus a pinhole mask grid. The ratio of the measured and known radius is a scale factor to account for changes in optical gain caused by atmospheric turbulence\cite{esposito2020}. The timing of this motion must be synchronized perfectly with exposures of the PWFS detector, and must also be precisely phased with the rest of the AO system (including the LGS WFS, OIWFS and ODGW, depending on the observing mode). These motions and timing features are illustrated in Figure~\ref{fig:waveforms}. Finally, the FSM must operate both at room temperature and the cold -30 C operating temperature of NFIRAOS\cite{densmore2020}, and provide good endurance (minimum 20 year service lifetime).

Trade studies during the NFIRAOS design phase led to the selection of a monocrystalline piezo actuator system, from Physik Instrument (PI), since more conventional polycrystalline devices would not meet the lifetime, stroke, and frequency requirements noted above. PI also claims excellent stability and hysteresis with similar piezo stages meaning that the FSM can function without sensor feedback.

In this paper we describe a prototype FSM based on the model P-915K925 stage and two-axis E-500/E-501 series piezo controller from PI\footnote{https://www.physikinstrumente.com/en/products/controllers-and-drivers/nanopositioning-piezo-controllers/e-500-e-501-modular-piezo-controller-601100/}, and characterize its performance in a series of experiments during which it is driven with a circular waveform:
\begin{itemize}
\item dynamic response, using driving waveforms at a range of frequencies,
\item linearity, by varying the radii of the waveform,
\item step response and measurements of mechanical delay,
\item temperature performance, repeating experiments both at room temperature and at -30\,C, and
\item absolute time synchronization.
\end{itemize}

%All but the final absolute timing test successfully met the stringent requirements of the PWFS. The timing difficulties that we encountered were caused by an integrated digital function generator (provided by PI as part of the same unit) that periodically missed externally-provided clock ticks, leading to erratic phase shifts. For the earlier tests the function generator was simply allowed to run on its own clock with no issue. 

%
% SETUP AND INITIAL CALIBRATION
%

\section{Experimental Setup and Initial Calibration}
\label{sec:setup}

\begin{figure}[ht]
  \begin{center}
    \includegraphics[width=\linewidth]{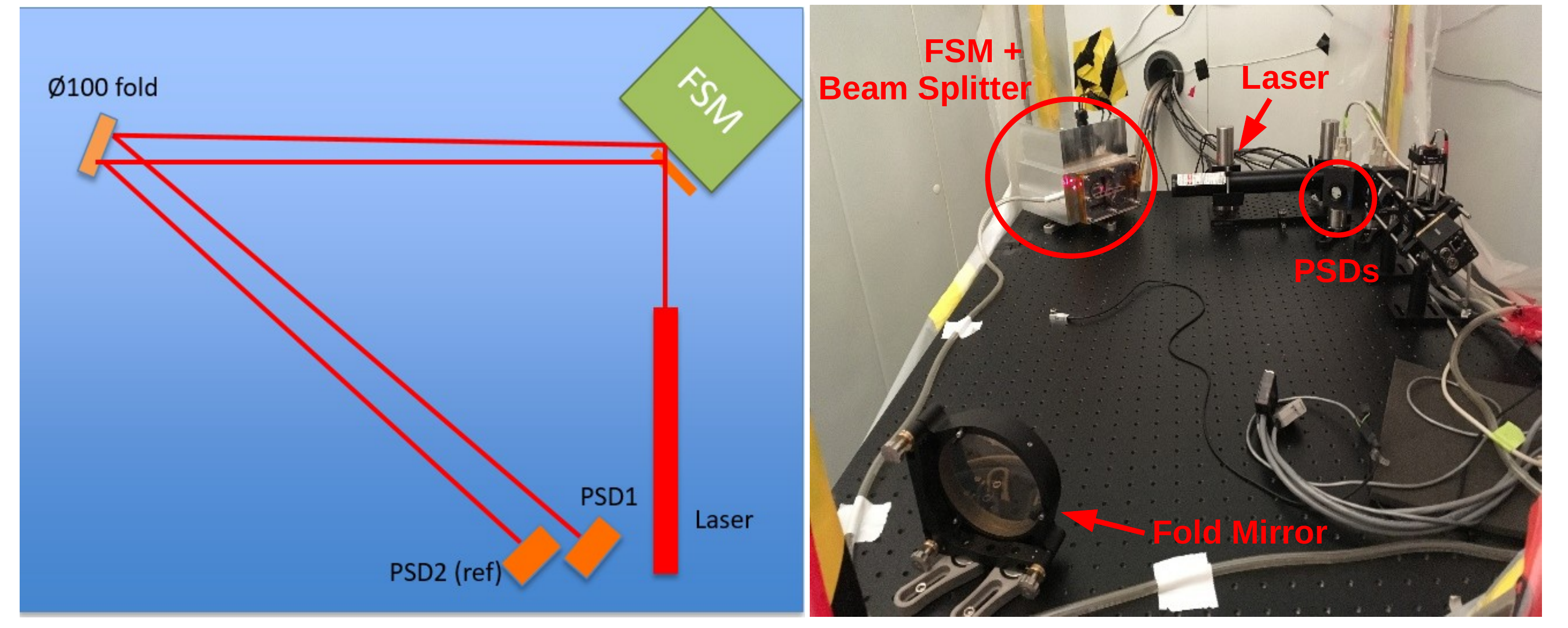}
  \end{center}
  \caption{\label{fig:setup}
  Optical bench setup, shown schematically on the left, and with the real devices on the right.}
\end{figure}

A laser, simulating the NGS beam, is directed to the FSM assembly which consists both of the FSM itself and a beam splitter, both mounted on a rigid block, so as to produce a second reference beam of light. The waveforms used to drive the FSM were provided by a PI E-518 Digital Interface and Function module, that was packaged by PI in the same chassis with the E-500 amplifier. The beams travel nearly parallel to one another, first to a common fold mirror to increase the level arm, and then to a pair of 2-axis high-speed position sensing devices (PSDs), one for each beam. Differential measurements can thus be performed to remove the effects of thermal drifts and vibrations on the table and other optical components. The PSDs produce voltages proportional to the locations of the laser spots in the range $\pm10$\,V for each axis. The layout of the optical bench is shown in Figure~\ref{fig:setup}. A Measurement Computing USB-1808X data acquisition unit (DAQ)\footnote{https://www.mccdaq.com/data-acquisition-and-control/simultaneous-daq/USB-1808-Series.aspx} was used to digitize these signals at up to 200\,kHz for later data analysis. Additional channels in the DAQ were used to concurrently digitize the external clock signal used to establish the absolute timing of the observations, and in a separate test, to sample the ouput voltages of the PI controller when not connected to the FSM (by means of a voltage divider, since the control voltages reach $\pm500$\,V). Both a GUI application and a lower-level software API are provided for interacting with the controller. For the initial calibration described in this section, vendor-supplied Windows GUIs were used to configure and operate both the PI Controller and DAQ.

Once the optical components had been mounted and aligned, the first task was to establish the correct waveform for each FSM axis to produce a circular pattern at the location of the PSDs. Due to the $45^\circ$ incidence angle between the laser and the FSM, an elliptical pattern is thus required. A linear transformation between commanded $T=(x,y)$ coordinates in actuator space, and measured $C = (a,b)$ coordinates on the PSDs, using an Interaction Matrix $\mathfrak{I}$ was established empirically:
\begin{gather*}
T_i = 
\begin{pmatrix}
  x_i \\
  y_i
\end{pmatrix}, \qquad \mathfrak{I} =
\begin{pmatrix}
  m_{11} & m_{12} \\
  m_{21} & m_{22}
\end{pmatrix}, \qquad C_i =
\begin{pmatrix}
  a_i \\
  b_i
\end{pmatrix} , \\[5pt]
T_i \mathfrak{I} = C_i .
\end{gather*}
The two axes of the stage were independently poked with both positive and negative values, and the resulting positions on the PSDs recorded. This procedure resulted in four sets of $T_i$ inputs and corresponding $C_i$ outputs, from which the coefficients of $\mathfrak{I}$ could be uniquely determined.

Multiplying each side of the equation by the inverse of the interaction matrix one can determine the commands, $T_i$, that would produce the desired coordinates on the PSDs, $C_i$,
\begin{gather*}
  T_i = C_i \mathfrak{I}^{-1} .
\end{gather*}

With this information in hand, output circular waveforms (sine and cosine waves) were converted into digitized input wavetable commands for each of the axes on the E-518 module. We initially used a free-running mode in which, once requested to start via a software command, the E-518 steps through the wavetable at a maximum rate of 25.0\,kHz using an internal clock, and repeats continuously until commanded to stop. The sample period is restricted to being an integer multiple of what the manual refers to as its internal ``servo period'' of 40\,$\mu$s. In order to simulate a waveform that matches the peak NFIRAOS requirement (800\,Hz), we settled on a 125-point wavetable containing four periods of the desired 800\,Hz waveform, and command the E-518 to step through it at the maximum 25.0\,kHz sample rate\footnote{This works out to $(1/800\,\mathrm{Hz})/(1/25000\,\mathrm{Hz}) = 31.25$ samples per period. The four periods were selected so that the waveform would fit into an integer number of samples, $N=125$. We also require four periods to include the dither signal.}.

Once the E-518 was commanded to start, the DAQ would record several seconds of data, but only data taken after one second were used to mitigate the impact of startup transients for the initial analysis.

Noting that the measurements used to derive the Interaction Matrix relied on the step response of the x- and y-stages (they were commanded to a position, and were held there during the PSD observation), it was known in advance that the dynamic response of the system to the fast-sampled waveforms would differ due to, e.g., the inertia of the stage, electronic delays, etc. Therefore, the output amplitudes of the resulting waveforms for each axis, $A$, and their phases, $\phi$, both with respect to the input demands, were calculated and applied to obtain the corrected waveforms, $T_c$:
\begin{gather*}
  T_c(t_i) =
  \begin{pmatrix}
    A_x x(t_i - \phi_x) \\
    A_y y(t_i - \phi_y)
  \end{pmatrix}
\end{gather*}

\begin{figure}[ht]
  \begin{center}
    \includegraphics[width=0.7\linewidth]{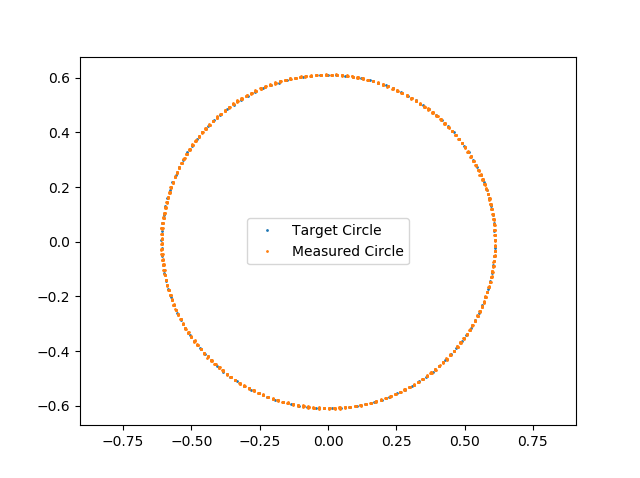}
  \end{center}
  \caption{\label{fig:optimization}
  Measured PSD positional data (yellow dots) compared with the target shape (blue dots) following optimization of the phase and amplitude of the wavetables to account for the dynamic response of the system. The two sets of points are nearly indistinguishable due to the accuracy of the control system.}
\end{figure}

Iterating this procedure a handful of times (i.e., re-running using the corrections from the previous iteration) demonstrated the ability to drive the stage in a highly circular pattern with the desired amplitude, as shown in Figure~\ref{fig:optimization}.

%
% STAGE CHARACTERIZATION
%

\section{Stage Characterization}

Having established a procedure for tuning the control waveform to achieve a particular output, we sought to answer several questions: (i) how does the dynamic response depend on the frequency of the driving waveform; (ii) how repeatably does the system behave; (iii) does the stage perform adequately at the -30\,C temperature of NFIRAOS;
(iv) how linear is the stage response; and (v) what is the startup (step) response. The tests described in the next subsections address these questions. Much of the data acquisition for these tests was orchestrated using Python scripts that communicate both with the DAQ and PI controller, using vendor-supplied APIs.

% DYNAMIC RESPONSE

\subsection{Dynamic response}

\begin{figure}[ht]
  \begin{center}
    \includegraphics[width=0.8\linewidth]{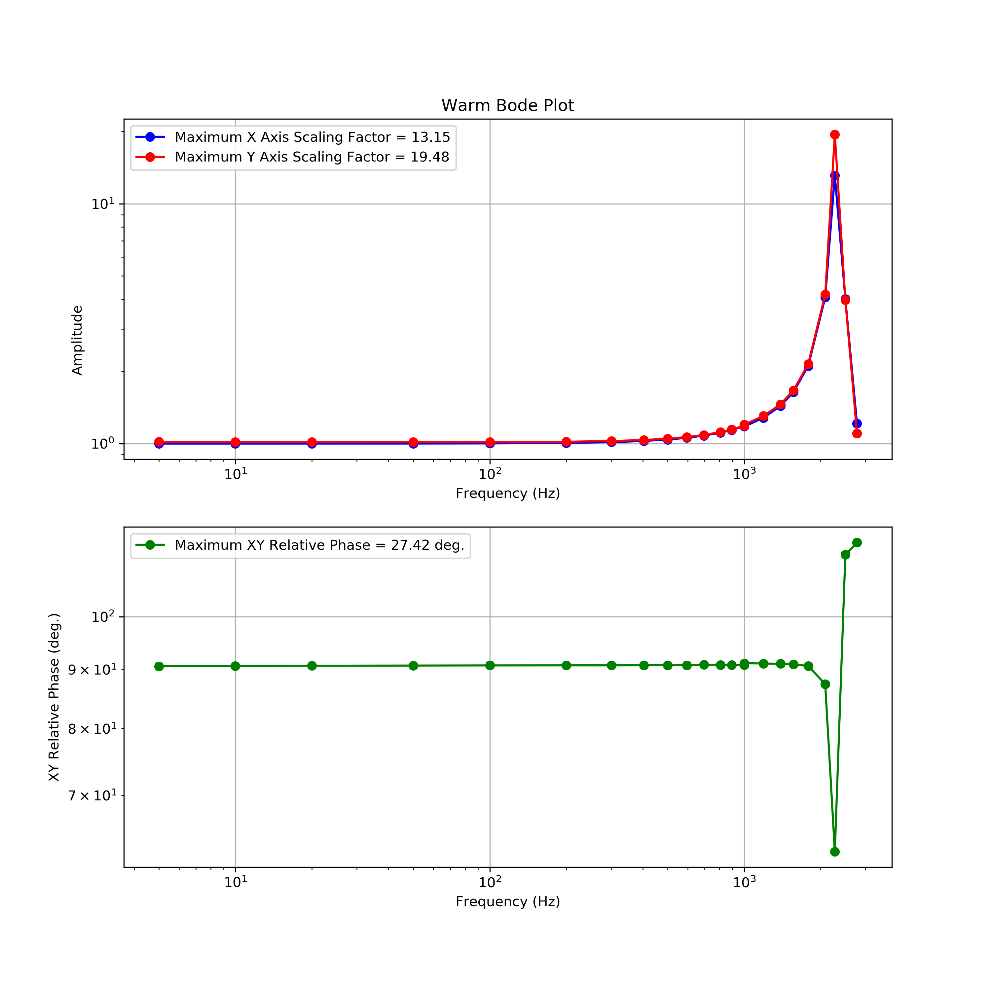}
  \end{center}
  \caption{\label{fig:warm_bode}
  Bode diagram for the FSM test system, measured at ambient temperature, at waveform frequencies ranging from 1\,Hz to 2.8\,kHz. The amplitude scale factors, $A$, and phase corrections, $\phi$, are determined iteratively.}
\end{figure}

To test the dynamic performance of the stage, we ran the stage with waveform frequencies ranging from 1\,Hz to 2.8\,kHz. In each case several seconds of data were acquired to ensure that the system had stabilized. The iterative procedure described previously was then used to determine the amplitude and phase shift fitting to several waveform periods of clean data.

Figure~\ref{fig:warm_bode} plots the corrections as a Bode diagram (nothing that ``amplitude'' in this plot is the reciprocal of the amplitude correction factor). At the low-frequency end the results match the steady-state response originally used to establish the interaction matrix; $A=1$, and $\phi=90$\,deg (noting that 90\,deg was the arbitrarily chosen zero-point for the phase in the this plot).

The amplitude then grows steadily toward a clear resonant peak near 2.4\,kHz. This result is consistent with PI's assertion that there are ``no resonant frequencies below 2.3\,kHz'', and is acceptable for the FSM requirement to run at 800\,Hz.

% REPEATABILITY

\subsection{Repeatability}

\begin{figure}[ht]
  \begin{center}
    \includegraphics[width=0.8\linewidth]{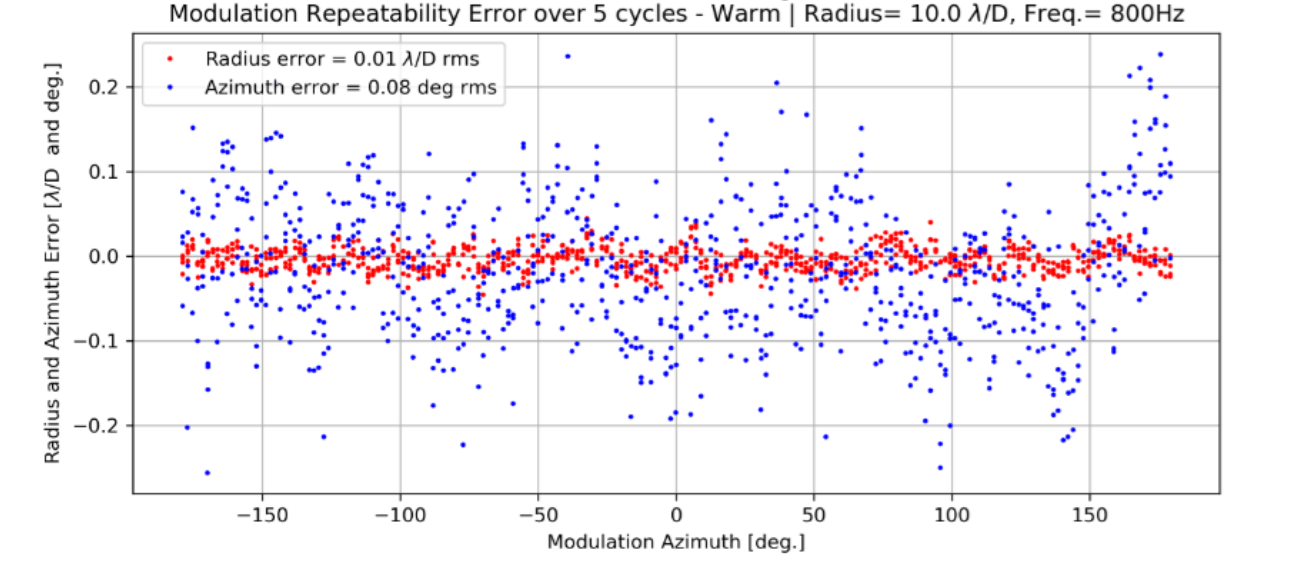}
  \end{center}
  \caption{\label{fig:repeatability}
  Repeatability shown as time-series errors for the radial and azimuthal coordinates over five periods at a driving frequency of 800\,Hz.}
\end{figure}

Repeatability of the stage performance was established by running the experiment over long periods (hours), acquiring data for periods of several seconds (since the DAQ is incapable of buffering data for more than 10\,s at the highest acquisition sample rate), and observing the error between the requested and measured waveforms. This experiment was repeated at a range of driving frequencies.

Figure~\ref{fig:repeatability} shows a representative example of the error signal in polar coordinates over five periods when driving the stage at 800\,Hz. The radial error (amplitude) scaled into units of $\lambda/\mathrm{D}$ has an RMS of 0.01, and the angular error (phase) has an RMS of 0.08\,deg. Both of errors are within the requirements for the FSM.

% LINEARITY

\subsection{Linearity}

\begin{figure}[ht]
  \begin{center}
    \includegraphics[width=0.8\linewidth]{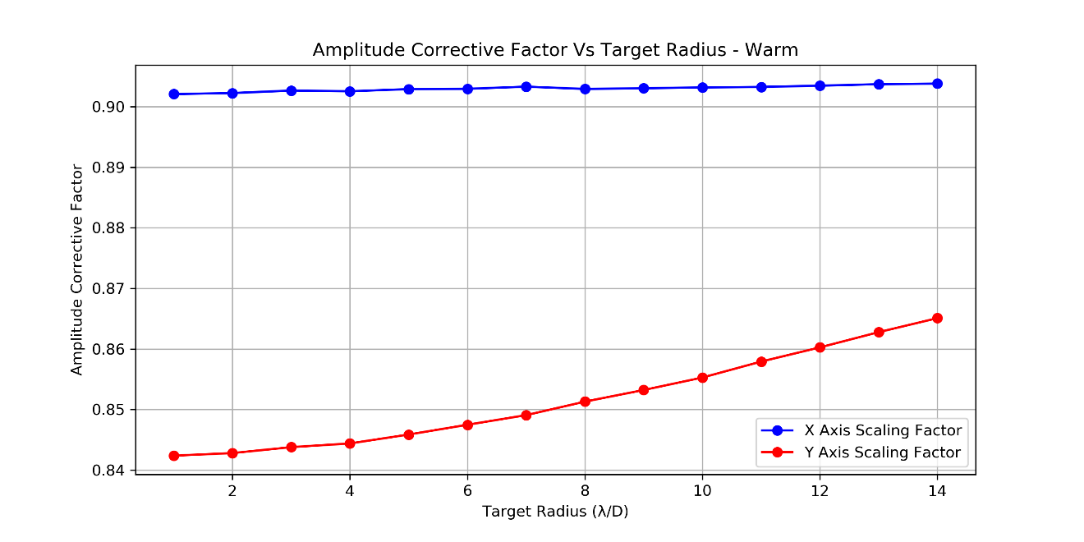}
  \end{center}
  \caption{\label{fig:linearity}
  Linearity tests in which the amplitude of the waveform is varied while the driving frequency is held at 800\,Hz. Optimized amplitude correction coefficients are shown for each axis.}
\end{figure}

Another area of interest is the linearity of the stage, i.e., does the amplitude correction factor vary as the target waveform amplitude is changed? To this end, data were collected for target amplitudes ranging from 1 to 14 in units of $\lambda/\mathrm{D}$, and again iteratively obtain the correction factors for each run.

Figure~\ref{fig:linearity} shows the full set of optimized amplitudes, for each axis. Note that the scale factor in X is approximately constant, though there is a clear variation in the correction factor in the Y-direction as the amplitude increases.

This observed asymmetric non-linearity could be due to the PSD, the drive electronics, or the actuator itself. Regardless of the cause, we have concluded that it will be possible to calibrate the system with a two-dimensional lookup table as a function of waveform radius and driving frequency.

% STARTUP RESPONSE

\subsection{Startup response}
\label{sec:startup}

\begin{figure}[ht]
  \begin{center}
    \includegraphics[width=\linewidth]{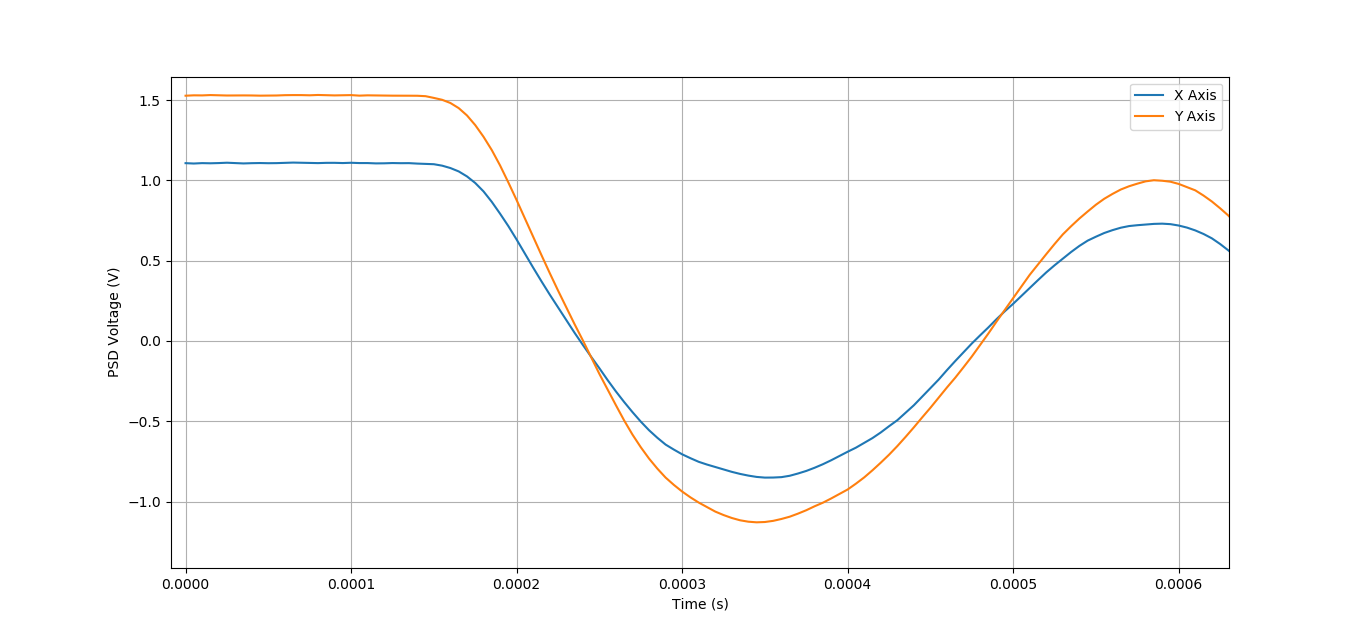}
  \end{center}
  \caption{\label{fig:startup_response}
  Startup response of the system. A simultaneous start trigger is sent to both the DAQ and PI controller ensure that they are synchronized.}
\end{figure}

Next we characterize the startup reponse of the system. This is accomplished by measuring the absolute electro-mechanical lag of the stage following the initial command to start motion.

In order to provide an absolute time reference, both the DAQ and PI Controller were placed into externally-triggered start modes. Once primed, both units await a shared trigger (via a tee) provided by a separate function generator (operated manually), and when received, the DAQ begins acquisition in parallel with the controller commencing playback of the driving waveform. The results are shown in Figure~\ref{fig:startup_response}, demonstrating an approxiumately 150\,$\mu$s lag between the start trigger and initial motion of the stage.

% COLD TESTS

\subsection{Cold tests}

In addition to the warm testing mentioned above, the FSM was further operated in a cold room at -30\,C to mimic the NFIRAOS environment. Repeating most of the previous tests, the stage was found to respond comparably in the cold. The main difference noted was that the amplitude scale factors increased slightly (for both axes), as did the frequency of the resonant peak. This behaviour is consistent with the idea that the piezo stage became stiffer at the lower temperature. These effects are easily characterized, and the increased frequency of the resonant peak is actually mildly beneficial.

%\begin{figure}[ht]
%  \begin{center}
%    \includegraphics[width=0.8\linewidth]{cold_bode.png}
%  \end{center}
%  \caption{\label{fig:cold_bode}
%  Bode diagram for the FSM test system, as in Figure~\ref{fig:warm_bode}, but at the -30\,C operating temperature of NFIRAOS}
%\end{figure}

%\textbf{...wasn't there also a long-term stability test? like periodic checks spread over a couple of days or something? ...}

%
% ABSOLUTE SYNCHRONIZATION
%

\section{Absolute synchronization}

The previous tests that the PI stage itself performs adequately for the purposes of NFIRAOS (the response can be characterized, and it appears to be stable over time). A final consideration that we wished to address was whether the waveforms output but the E-518 could be synchronized with an external trigger. The unit offers two basic modes for making using of an external trigger that might support this operation: (i) as a start trigger (i.e., the unit free-runs once started) as used in Section~\ref{sec:startup}; and (ii) as a clock signal. Option (i) was immediately ruled-out as the internal clock of the unit (used to step through the wavetable) cannot be tied to an external reference (such as the Precision Time Protocal for computers on a network), meaning that while one could start it at the correct absolute time, it would slowly drift with respect to the rest of the AO system. Option (ii) looked promising since each clock ``tick'' is used to step to the next sample in the wavetable. As long as the waveform playback is started at the correct time, and the clock signal detection is reliable (i.e., no ticks are dropped), the unit could be expected to operate in phase with the rest of the AO system.

\begin{figure}[ht]
  \begin{center}
    \includegraphics[width=\linewidth]{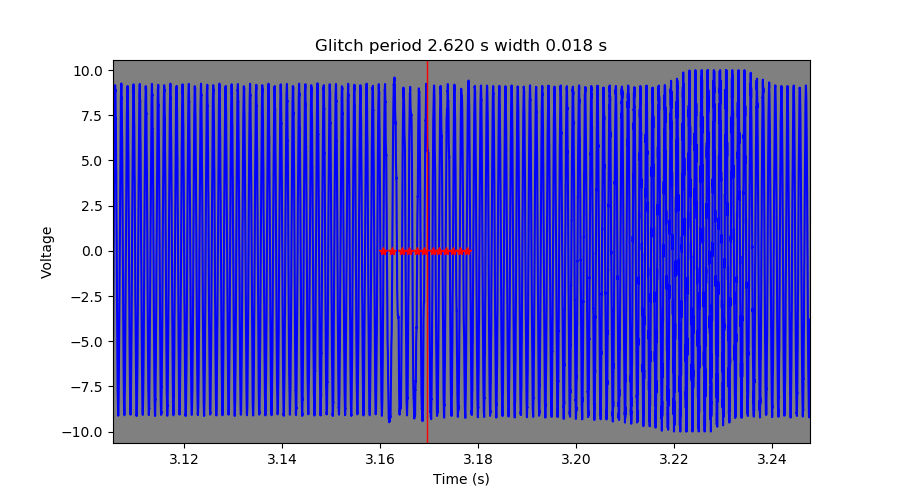}
  \end{center}
  \caption{\label{fig:glitch}
  Output voltage for one of the controller axes over time, exhibiting a timing ``glitch'' while using an external 12.5\,kHz lock signal to step through the wavetable (producing an 800\,Hz waveform). The output is highly distorted in a burst lasting approximately 0.018\,s after it settles to a random phase. These bursts were observed to occur periodically (once every $\sim$2.62\,s).}
\end{figure}

%\begin{figure}[ht]
%  \begin{center}
%    \includegraphics[width=0.8\linewidth]{glitch_zoomedOut.png}
%  \end{center}
%  \caption{\label{fig:glitch_zoomedOut}
%  Glitch zoomed out.}
%\end{figure}

Unfortunately, testing with the 800\,Hz waveform immediately revealed a strange behaviour: after initially running as expected for several seconds, there would be brief periods during which the unit would appear to miss several clock tickets. Shortly thereafter it would resume functioning normally, but now with a different (and random) phase. We experimented a great deal with the signal generator used to produce the clock signal, including different clock waveforms (i.e., ramps with various rise and fall times, and different amplitudes), and running the unit at a range of frequencies. Unfortunately we were never able to resolve the issue, though our working theory is that for small fraction of the internal 40\,$\mu$s servo period the E-518 suffers a ``dead window'' during which it is incapable of sensing the clock signal. Since the internal clock of the E-518 and the external trigger generator will slowly drift with respect to one another, one might expect a burst of clock signals to land repeatedly within this dead period if they occur with a frequency that is an integer multiple of the 40\,$\mu$s servo period (i.e., until the clock drifts sufficiently for them to again land outside of the dead window). Furthermore, if the external clock runs at a \emph{non-integer multiple} of the servo period, the times at which clock ticks are missed would tend to be more haphazard (i.e., one tick might land in the dead window, with subsequent ticks landing far outside of it).

Figure~\ref{fig:glitch} shows the output voltage from the controller for one of the axes (passed through a voltage divider to enable digitization with the DAQ). For this test we again use a 125-sample wavetable, but this time with 8 periods of our desired 800\,Hz waveform, with the clock running at 12.5\,kHz (15.625 sampels per period). Running with the slower clock enables us to see the shape of the digitized clock signal more clearly with the DAQ, but the results are qualitatively similar to those obtained with the higher-clock rate waveform described earlier. A Python script was written to step through the captured time series and identify the boundaries of each period from the zero-crossings. It was then possibly to identify outlier periods with respect to the median, which are shown with the red stars in the figure. Clearly the anomolous periods occur in a localized burst over an interval of roughly 0.018\,s. Examing the same data on longer time-scales it was found that these bursts occur regularly, with a period of about 2.620\,s. If we assume that this period corresponds to the controller clock drifting with respect to the external function generator by a 40\,$\mu$s servo period, that works out to a drift rate of $40\,\mu\mathrm{s} / 2.620\,\mathrm{s} = 15.3\,\mu\mathrm{s}/\mathrm{s}$, which is plausible. Multiplying the observed burst interval by this drift rate gives us an estimate of the width of this hypothesized dead window, $0.018\,s \times 15.3\,\mu\mathrm{s}/\mathrm{s} = 0.275\,\mu\mathrm{s}$. Re-running this test on different days showed qualitatively similar behaviour, though the period of the bursts varied slightly (e.g., sometimes slightly above 3\,s, sometimes below). Again, this is consistent with the hypothesis that the problem is correlated with clock drift (with a rate that, itself, can also drift).

Finally, we repeated this test using a number of other external clock rates that do not correspond to integer multiples of the servo period. Though not shown here, we found that the anomolous periods would typically not occur in bursts as described above, lending further support to our theory.

Unfortunately we were not able to remedy this issue with the manufacturer, and will have to consider other methods for establishing absolute time synchronization.

\section{Summary and future work}

The NFIRAOS team is satisfied that the selected monocrystalline piezo stage and E-500 controller from PI meet the basic requirements of the FSM. The mirror was shown to accurately reproduce the desired circular waveform at up to 800\,Hz, both at ambient temperature and at the -30\,C temperature of NFIRAOS. A procedure was established for calibrating the stage (coordinate transformations, amplitude and phase corrections), and the prototype exhibited good long-term stability. The only outstanding issue is synchronizing its motion with an external trigger; unfortunately we were unable to use the integrated PI E-518 module in external clocking mode due to it periodically failing to detect some clock ticks. Both our team and our contacts at PI agree that we will have to bypass the integrated E-518 digital waveform module and directly feed the E-500 amplifiers with analog signals from a custom function generator.

\acknowledgments

The TMT Project gratefully acknowledges the support of the TMT collaborating institutions.  They are the California Institute of Technology, the University of California, the National Astronomical Observatory of Japan, the National Astronomical Observatories of China and their consortium partners, the Department of Science and Technology of India and their supported institutes, and the National Research Council of Canada.  This work was supported as well by the Gordon and Betty Moore Foundation, the Canada Foundation for Innovation, the Ontario Ministry of Research and Innovation, the Natural Sciences and Engineering Research Council of Canada, the British Columbia Knowledge Development Fund, the Association of Canadian Universities for Research in Astronomy (ACURA) , the Association of Universities for Research in Astronomy (AURA), the U.S. National Science Foundation, the National Institutes of Natural Sciences of Japan, and the Department of Atomic Energy of India.

% References
\bibliography{refs} % bibliography data in refs.bib
\bibliographystyle{spiebib} % makes bibtex use spiebib.bst

\end{document}